\documentclass[twoside,12pt]{report}
\usepackage{graphicx}
\usepackage{amssymb, amsmath, amsxtra}
\usepackage{epsfig,subfigure}

\usepackage{latexsym}
\usepackage{bm}

\begin{document}
 \begin{center}
 {\Large \bf Heat conduction in anisotropic media.}\\[.6ex] {\large \bf Nonlinear self-adjointness and conservation laws}\\[2ex]
  Nail H.~Ibragimov\\
 \textit{Laboratory "Group analysis of mathematical models\\ in natural and engineering sciences",}\\
  \textit{Ufa State Aviation Technical University,\\
   450 000 Ufa, Russia\\[.1ex]
   and Research centre ALGA, \\Department of Mathematics and Science,\\
   Blekinge Institute of Technology,\\ SE-371 79 Karlskrona, Sweden}\\[1ex]
  Elena D. Avdonina\\
 \textit{Laboratory "Group analysis of mathematical models\\ in natural and engineering sciences",}\\
  \textit{Ufa State Aviation Technical University,\\
   450 000 Ufa, Russia}

 \end{center}

 \begin{small}
 \noindent
{\bf Abstract}\\
Nonlinear self-adjointness  of the anisotropic nonlinear heat
equation is investigated. Mathematical models of heat conduction in
anisotropic media with a source are considered and a class of
self-adjoint models is identified.
Conservation laws corresponding to the symmetries of the equations in question  are computed.\\


 \noindent
 \textit{AMS classification numbers}:
 35C99, 70S10, 70G65
 \end{small}
 \hfill

 \begin{table}[b]
  \begin{tabular}{l}\hline
 \copyright ~2012 N.H. Ibragimov and E.D. Avdonina.\\
  \end{tabular}
 \end{table}

 \tableofcontents
  \newpage

 \section{Introduction}
 \label{anh-int}
 \setcounter{equation}{0}

 \subsection{Formulation of the problem}
 \label{anh-int.1}

 The nonlinear second-order evolution equation
 \begin{equation}
 \label{anh-int:eq1}
  u_t = (f(u) u_x)_x + (g(u) u_y)_y + (h(u) u_z)_z
 \end{equation}
 describes the heat conduction in anisotropic materials whose physical characteristics such as the thermal conductivity
 are affected by the temperature. Physical applications and interesting
 mathematical properties of Eq. (\ref{anh-int:eq1}) and of its
 extension to the case of existence of an external source $q(u),$
 \begin{equation}
 \label{anh-int:eq2}
  u_t = (f(u) u_x)_x + (g(u) u_y)_y + (h(u) u_z)_z + q(u)
 \end{equation}
  are discussed
 in \cite{gdeks88} (see also \cite{ibr94}, Section 10.9).

 The functions $f(u), g(u), h(u)$ are positive according to their physical
 meaning. So, we consider Eqs. (\ref{anh-int:eq1}) and
 (\ref{anh-int:eq2}) with arbitrary positive coefficients  $f(u), g(u), h(u).$
 Furthermore, we will assume that these coefficients are linearly independent and that none of them
 is constant, i.e.
 \begin{equation}
 \label{anh-int:eq3}
  f'(u) \not= 0, \quad g'(u) \not= 0, \quad h'(u) \not= 0.
 \end{equation}

 Note, that Eq. (\ref{anh-int:eq1}) has a conservation form whereas Eq. (\ref{anh-int:eq2})
 with $q(u) \not= 0$ does not have such form. A conservation form is useful in many respects, e.g.
 in qualitative and numerical analysis. Moreover, possibility of different conservation forms can be helpful.
 Therefore we will construct various conservation laws for Eq. (\ref{anh-int:eq1}) using the method
 of nonlinear self-adjointness \cite{ibr10a} and investigate
  the question on existence of conservation laws for  Eq.
  (\ref{anh-int:eq2}) with specific values of the source term $q(u) \not= 0.$

 \subsection{Nonlinear self-adjointness}
 \label{anh-int.2}

 Recall the definition of nonlinear self-adjointness.
 Let us consider a second-order partial differential equation
 \begin{equation}
 \label{anh-int.2:eq.1}
 F\big(x, u, u_{(1)}, u_{(2)}\big) = 0,
  \end{equation}
 where $u$ is  the dependent variable,
  $u_{(1)}$ and $u_{(2)}$ are the sets of
 the first-order partial derivatives $u_i$ and the second-order derivatives $u_{ij}$ of $u$ with respect to the independent
 variables $x = (x^1, \ldots, x^n).$ The \textit{adjoint equation} to Eq. (\ref{anh-int.2:eq.1})
  is
 \begin{equation}
 \label{anh-int.2:eq.2}
 F^*\big(x, u, v, u_{(1)}, v_{(1)}, u_{(2)}, v_{(2)}\big)
  = 0,
 \end{equation}
 where $F^*$ is defined by
 \begin{equation}
 \label{anh-int.2:eq.3}
 F^*\big(x, u, v, u_{(1)}, v_{(1)}, u_{(2)}, v_{(2)}\big)
 = \frac{\delta (v F)}{\delta u}\,\cdot
 \end{equation}
 Here  $v$ is a new dependent variable and $v_{(1)}, v_{(2)}$
 are the sets of its partial derivatives. Furthermore, ${\delta (v F)}/{\delta u}$
 denotes the variational derivative of $vF:$
 $$
 \frac{\delta (v F)}{\delta u} =
 \frac{\partial (vF)}{\partial u} -
 D_i \left(\frac{\partial (vF)}{\partial u_i} \right)
 + D_i D_k \left(\frac{\partial (vF)}{\partial u_{ik}} \right)
 - \cdots\,,
 $$
 where the total differentiations are extended to the new dependent variable
 $v:$
 \begin{equation}
 \label{anh-int.2:eq.4}
  D_i = \frac{\partial}{\partial x^i} +
 u_{i}\frac{\partial}{\partial u} +
 v_{i}\frac{\partial}{\partial v} +
 u_{ij}\frac{\partial}{\partial u_j} +
 v_{ij}\frac{\partial}{\partial v_j} +
 \cdots\,.
 \end{equation}

 Eq. (\ref{anh-int.2:eq.1})  is
 said to be \textit{nonlinearly self-adjoint}  \cite{ibr10a} if
 the adjoint equation (\ref{anh-int.2:eq.2})  is satisfied for all solutions $u$
of the original equation (\ref{anh-int.2:eq.1}) upon a
substitution\footnote{In general, the substitution
(\ref{anh-int.2:eq.5}) can be of the form $v = \varphi\big(x, u,
u_{(1)}\big).$}
  \begin{equation}
  \label{anh-int.2:eq.5}
 v = \varphi(x, u),\quad \varphi \not= 0.
 \end{equation}
The condition that the function $\varphi$ that does not vanish is
significant. The condition for the nonlinear self-adjointness can be
written in the form
   \begin{equation}
  \label{anh-int.2:eq.6}
 F^*\big(x, u, \varphi, u_{(1)}, \varphi_{(1)}, u_{(2)}, \varphi_{(2)}\big) = \lambda\, F\big(x, u,
 u_{(1)}, u_{(2)}\big),
  \end{equation}
  where  $\lambda = \lambda (x, u, u_{(1)}, \ldots)$ is an undetermined variable coefficient and $\varphi_{(1)},
  \varphi_{(2)}$ denote the derivatives of the function $\varphi.$
  E.g. $\varphi_{(1)}$ is the set of the first-order total
  derivatives
  $$
 D_i (\varphi) = \frac{\partial \varphi(x, u)}{\partial x^i}
 + u_i \frac{\partial \varphi(x, u)}{\partial u}\,, \quad i = 1,
 \ldots, n.
   $$
   Eq. (\ref{anh-int.2:eq.6}) should be satisfied identically in all
   variables $x, u, u_{(1)}, u_{(2)}.$

 \subsection{Conserved vector associated with symmetries}
 \label{anh-int.3}

 The general result  on construction of conserved vectors
 associated with symmetries of nonlinearly self-adjoint equations
 demonstrated in \cite{ibr10a} leads to the following statement for the second-order equation
 (\ref{anh-int.2:eq.1}).

  Let (\ref{anh-int.2:eq.1}) be nonlinearly self-adjoint and admit a one-parameter point transformation group with the
  generator
 \begin{equation}
 \label{anh-int.3:eq.1}
 X = \xi^i(x, u) \frac{\partial}{\partial x^i} +
 \eta(x,u) \frac{\partial}{\partial u}\,\cdot
 \end{equation}
 Then the vector
 \begin{equation}
 \label{anh-int.3:eq.2}
  C^i  = W \left[\frac{\partial {\cal L}}{\partial u_i} -
  D_j \left(\frac{\partial {\cal L}}{\partial u_{ij}}\right)\right]
 + D_j(W) \frac{\partial {\cal L}}{\partial u_{ij}}\,, \quad i = 1,
 \ldots, n,
 \end{equation}
 is a \textit{conserved vector} for Eq. (\ref{anh-int.2:eq.1}), i.e. satisfies the conservation equation
 \begin{equation}
 \label{anh-int.3:eq.3}
  \left[D_i(C^i)\right]_{(\ref{anh-int.2:eq.1})}  = 0.
 \end{equation}
 Here
 \begin{equation}
 \label{anh-int.3:eq.4}
 W = \eta  - \xi^j u_j
 \end{equation}
 and ${\cal L}$ is the \textit{formal Lagrangian}  for Eq. (\ref{anh-int.2:eq.1}) given by
 \begin{equation}
 \label{anh-int.3:eq.5}
 {\cal L} = v F.
 \end{equation}
  It is assumed that the variable $v$ and its derivatives
  are eliminated from the right-hand side of  Eq. (\ref{anh-int.3:eq.2}) by using
  the substitution (\ref{anh-int.2:eq.5}), where the function $\varphi (x, u)$ is
  found by solving the nonlinear self-adjointness condition (\ref{anh-int.2:eq.6}).

 \section{Investigation of nonlinear self-adjointness}
 \label{anh-1}
 \setcounter{equation}{0}

 \subsection{Substitution (\ref{anh-int.2:eq.5}) for Equation
 (\ref{anh-int:eq1})}
  \label{anh-1.subst}

 Since Eq. (\ref{anh-int:eq1}) has the conservation form
 (\ref{anh-int.3:eq.3}), it is nonlinearly self-adjoint by Theorem 8.1
 from \cite{ibr10a}. Let us find the corresponding substitution
 (\ref{anh-int.2:eq.5}).

 We write Eq. (\ref{anh-int:eq1}) in the form
 (\ref{anh-int.2:eq.1}):
 \begin{equation}
 \label{anh-1:eq1}
 F\equiv -u_t +f(u) u_{xx} +g(u)u_{yy}+h(u)u_{zz} + f'(u) u_x^2 +g'(u) u_y^2 +h'(u)
 u_z^2=0,
 \end{equation}
insert the expression for $F$ in (\ref{anh-int.2:eq.3}) and after
simple calculations obtain the following adjoint equation
(\ref{anh-int.2:eq.2}) to Eq. (\ref{anh-int:eq1}):
 \begin{equation}
 \label{anh-1:eq2}
  F^*\equiv  v_t+f(u) v_{xx} +g(u) v_{yy} +  h(u) v_{zz}=0.
  \end{equation}

 In our case the substitution (\ref{anh-int.2:eq.5}) has the form
 \begin{equation}
 \label{anh-1:eq3}
 v = \varphi (t, x, y, z,  u).
 \end{equation}
 Its derivatives are written
 \begin{equation}
 \label{anh-1:eq4}
 \begin{split}
 & v_t \equiv D_t(\varphi) = \varphi_u u_t + \varphi_t,
 \quad v_x \equiv D_x(\varphi) = \varphi_u u_x + \varphi_x, \\[1ex]
 & v_y \equiv D_y(\varphi) = \varphi_u u_y + \varphi_y,
 \quad v_z \equiv D_z(\varphi) = \varphi_u u_z + \varphi_z,\\[1ex]
 & v_{xx} \equiv D_x^2(\varphi) = \varphi_u u_{xx} + \varphi_{uu} u_x^2 + 2 \varphi_{xu} u_x + \varphi_{xx},\\[1ex]
 & v_{yy} \equiv D_y^2(\varphi) = \varphi_u u_{yy} + \varphi_{uu} u_y^2 + 2 \varphi_{yu} u_y + \varphi_{yy},\\[1ex]
 & v_{zz} \equiv D_z^2(\varphi) = \varphi_u u_{zz} + \varphi_{uu} u_z^2 + 2 \varphi_{zu} u_z +
 \varphi_{zz}.
 \end{split}
 \end{equation}
 Now we take the nonlinear self-adjointness condition
 (\ref{anh-int.2:eq.6}), where $F$ and $F^*$ are given by (\ref{anh-1:eq1}) and (\ref{anh-1:eq2}),
 respectively:
 \begin{align}
 \label{anh-1:eq5}
 & v_t+f(u) v_{xx} +g(u) v_{yy} +  h(u) v_{zz} \\[1ex]
  = & \lambda\, \left[-u_t +f(u) u_{xx} +g(u)u_{yy}+h(u)u_{zz} + f'(u) u_x^2 +g'(u) u_y^2 +h'(u)
 u_z^2\right]\notag
 \end{align}
 where the corresponding derivatives of $v$ in the left-hand side should be replaced with their expressions (\ref{anh-1:eq4}).
  First we compare the coefficients for $u_t$ in
 both sides of Eq. (\ref{anh-1:eq5}) and obtain
 $$
 \lambda = - \varphi_u.
 $$
 Then the coefficients for $u_{xx}, u_{yy}, u_{zz}$ yield:
 $$
 f(u) \varphi_u = - f(u) \varphi_u, \quad g(u) \varphi_u = - g(u) \varphi_u, \quad
 h(u) \varphi_u = - h(u) \varphi_u.
 $$
 By our assumption, the functions $f(u), g(u), h(u)$  do not vanish. Therefore the above
 equations yield that $\varphi_u = 0.$ Hence, $\varphi = \varphi (t, x, y, z)$
 and therefore
 $$
 \lambda = 0, \quad v_t = \varphi_t, \quad v_{xx} = \varphi_{xx},
 \quad v_{yy} = \varphi_{yy}, \quad v_{zz} = \varphi_{zz}.
 $$
 Then Eq. (\ref{anh-1:eq5}) becomes
 \begin{equation}
 \label{anh-1:eq6}
 \varphi_t +f(u) \varphi_{xx} +g(u) \varphi_{yy} +  h(u) \varphi_{zz} = 0.
 \end{equation}
 Since  $f(u), g(u), h(u)$ are linearly independent
and obey the conditions (\ref{anh-int:eq3}), whereas $\varphi$ does
not depend on $u,$ Eq. (\ref{anh-1:eq6}) yields
 \begin{equation}
 \label{anh-1:eq7}
 \varphi_t = 0,\quad \varphi_{xx} = 0, \quad \varphi_{yy} = 0, \quad \varphi_{zz} = 0.
 \end{equation}
 The general solution of Eqs. (\ref{anh-1:eq7}) is given by
 $$
 \varphi = a_1\,xyz + a_2\,xy + a_3\,xz + a_4\,y z + a_5\,x + a_6\,y + a_7\,z + a_8
 $$
 with arbitrary constant coefficients $a_1,\ldots a_8.$ This proves the
 following.\\[1ex]
 \textbf{Proposition 2.1.}
 Eq. (\ref{anh-int:eq1}) satisfies the nonlinear self-adjointness condition
 (\ref{anh-int.2:eq.6}) with the substitution (\ref{anh-1:eq3}) of the form
 \begin{equation}
 \label{anh-1:eq8}
 v = a_1\,xyz + a_2\,xy + a_3\,xz + a_4\,y z + a_5\,x + a_6\,y + a_7\,z +
 a_8.
 \end{equation}

 \subsection{Two-dimensional equation with a source}
  \label{anh-1.source}

 Let us consider Eq. (\ref{anh-int:eq2}), for the sake of simplicity,  in the case of two spatial variables
 $x, y:$
 \begin{equation}
 \label{anh-int:eq9}
  u_t = (f(u) u_x)_x + (g(u) u_y)_y + q(u).
 \end{equation}
 The adjoint equation has the form
  \begin{equation}
 \label{anh-1:eq10}
  F^*\equiv  v_t+f(u) v_{xx} +g(u) v_{yy} +  q'(u) v=0.
  \end{equation}

 Repeating the calculations of Section \ref{anh-1.subst} we obtain
 the following equation for the nonlinear self-adjointness of Eq.
(\ref {anh-int:eq9}) (compare with Eq. (\ref{anh-1:eq6})):
 \begin{equation}
 \label{anh-1:eq11}
 \varphi_t +f(u) \varphi_{xx} +g(u) \varphi_{yy} +  q'(u) \varphi = 0.
 \end{equation}
 If $f(u), g(u)$ and $q(u)$ are arbitrary functions, Eq.
 (\ref{anh-1:eq11}) yields (compare with Eqs. (\ref{anh-1:eq7})):
 \begin{equation}
 \label{anh-1:eq12}
 \varphi_t = 0,\quad \varphi_{xx} = 0, \quad \varphi_{yy} = 0, \quad \varphi = 0.
 \end{equation}
 These equations show that a substitution of the form (\ref{anh-int.2:eq.5}) does not
 exist.
 Indeed, the last equation in (\ref{anh-1:eq12}) contradicts
 the condition $\varphi \not= 0.$
 Hence, Eq. (\ref {anh-int:eq9}) with the arbitrary source $q(u)$ is not
 nonlinearly self-adjointness with the substitution of the form
 (\ref{anh-int.2:eq.5}).

 However, Eq. (\ref {anh-int:eq9}) with sources of particular forms can be nonlinearly
 self-adjoint. For example, let
 \begin{equation}
 \label{anh-1:eq13}
 q'(u) = r f(u),\quad r = {\rm const.}
 \end{equation}
 Then Eq. (\ref{anh-1:eq11}) becomes
 $$
 \varphi_t +f(u) [\varphi_{xx} + r \varphi] + g(u) \varphi_{yy}  = 0
 $$
 and yields (compare with Eqs. (\ref{anh-1:eq12})):
 \begin{equation}
 \label{anh-1:eq14}
 \varphi_t = 0, \quad \varphi_{yy} = 0, \quad \varphi_{xx} + r \varphi = 0.
 \end{equation}
 The solution to Eqs. (\ref{anh-1:eq14}) has the form
 \begin{equation}
 \label{anh-1:eq15}
 \varphi = a(x) y + b(x),
 \end{equation}
 where $a(x)$ and $b(x)$ arbitrary solutions of the linear second-order ODE
 \begin{equation}
 \label{anh-1:eq16}
  w'' + r w = 0.
 \end{equation}

 Eq. (\ref{anh-1:eq13}) shows that the source strength increases
 together with the temperature, i.e. $q'(u) >0,$ if $r = \omega^2 > 0,$ and
 decreases, $q'(u) < 0,$ if $r = - \delta^2 < 0.$ Having this in mind and denoting
 \begin{equation}
 \label{anh-1:eq17}
 {\cal F}(u) = \int f(u) du
 \end{equation}
 we consider two particular forms of Eq. (\ref {anh-int:eq9}):
 \begin{equation}
 \label{anh-int:eq18}
  u_t = (f(u) u_x)_x + (g(u) u_y)_y + \omega^2 {\cal F}(u), \quad \omega =
  {\rm const.,}
 \end{equation}
 and
 \begin{equation}
 \label{anh-int:eq19}
  u_t = (f(u) u_x)_x + (g(u) u_y)_y - \delta^2 {\cal F}(u), \quad \delta =
  {\rm const.}
 \end{equation}

 In the case (\ref{anh-int:eq18}) Eq. (\ref{anh-1:eq16}) is
 written
 $$
  w'' + \omega^2 w = 0
 $$
 and yields
 $$
  w = C_1 \cos (\omega x) + C_2 \sin (\omega x).
 $$
 Hence
 $$
 a(x) = A_1 \cos (\omega x) + A_2 \sin (\omega x), \quad
 b(x) = B_1 \cos (\omega x) + B_2 \sin (\omega x)
 $$
 with arbitrary constants $A_1, A_2, B_1, B_2.$ We  we substitute these expressions in Eq. (\ref{anh-1:eq15})
 and arrive at the following statement.\\[1ex]
 \textbf{Proposition 2.2.}
 Eq. (\ref{anh-int:eq18}) satisfies the nonlinear self-adjointness condition
 (\ref{anh-int.2:eq.6}) with the substitution (\ref{anh-1:eq3}) of the form
 \begin{equation}
 \label{anh-1:eq20}
 v = \left(A_1 y + B_1\right) \cos (\omega x) + \left(A_2 y + B_2\right) \sin (\omega x).
 \end{equation}

In the case (\ref{anh-int:eq19}) Eq. (\ref{anh-1:eq16}) is
 written
 $$
  w'' - \delta^2 w = 0
 $$
 and yields
 $$
  w = C_1 {\rm e}^{\delta x} + C_2 {\rm e}^{-\delta x}.
 $$
 Proceeding  as above we arrive at the following statement.\\[1ex]
 \textbf{Proposition 2.3.}
 Eq. (\ref{anh-int:eq19}) satisfies the nonlinear self-adjointness condition
 (\ref{anh-int.2:eq.6}) with the substitution (\ref{anh-1:eq3}) of the form
 \begin{equation}
 \label{anh-1:eq21}
 v = \left(A_1 y + B_1\right) {\rm e}^{\delta x} + \left(A_2 y + B_2\right) {\rm e}^{-\delta x}.
 \end{equation}

 \subsection{Remark on materials with specific anisotropy}
  \label{anh-1.rem}

The situation is different if the conditions (\ref{anh-int:eq3}) are
not satisfied. Let, e.g. $g(u)$ be a positive constant, $g = k.$
Then Eq. (\ref{anh-int:eq1}) has the form
 \begin{equation}
 \label{anh-1:eq22}
 u_t = (f(u) u_x)_x + k u_{yy} + (h(u) u_z)_z.
 \end{equation}
In this case Eqs. (\ref{anh-1:eq7}) are replaced by the following
equations:
 \begin{equation}
 \label{anh-1:eq23}
 \varphi_t + k \varphi_{yy} = 0, \quad \varphi_{xx} = 0, \quad \varphi_{zz} = 0.
 \end{equation}
 The second and third equations of the system (\ref{anh-1:eq23})
 yield
 $$
\varphi = \alpha (t, y) xz + \beta (t, y) x + \gamma(t, y) z +
\sigma(t, y).
 $$
 The first equation
 (\ref{anh-1:eq23}) shows that $\alpha (t, y), \beta (t, y), \gamma(t,
 y)$ and $\sigma(t, y)$ solve the adjoint equation
 \begin{equation}
 \label{anh-1:eq24}
 v_t + k v_{yy} = 0
 \end{equation}
 to the linear heat equation
 \begin{equation}
 \label{anh-1:eq25}
 u_t - k u_{yy} = 0.
 \end{equation}
 Thus, we have demonstrated the
 following statement.\\[1ex]
 \textbf{Proposition 2.4.}
 Eq. (\ref{anh-1:eq22}) satisfies the nonlinear self-adjointness condition
 (\ref{anh-int.2:eq.6}) with the substitution (\ref{anh-1:eq3}) of the form
 \begin{equation}
 \label{anh-1:eq26}
 v = \alpha (t, y) xz + \beta (t, y) x + \gamma(t, y) z +
\sigma(t, y),
 \end{equation}
 where $\alpha (t, y), \beta (t, y), \gamma(t,
 y)$ and $\sigma(t, y)$ are any solutions of the adjoint equation (\ref{anh-1:eq24})
 to the linear heat equation (\ref{anh-1:eq25}).

 Combining Propositions 2.2 and 2.3 with Proposition 2.4, we obtain
 the following statements.\\[1ex]
 \textbf{Proposition 2.5.}
 The equation
 \begin{equation}
 \label{anh-int:eq27}
  u_t = (f(u) u_x)_x + k u_{yy} + \omega^2 {\cal F}(u), \quad f(u) = {\cal F}'(u),
 \end{equation}
  satisfies the nonlinear self-adjointness condition
 (\ref{anh-int.2:eq.6}) with the substitution (\ref{anh-1:eq3}) of the form
 \begin{equation}
 \label{anh-1:eq26}
 v = \alpha (t, y) \cos (\omega x) + \beta (t, y) \sin (\omega x),
 \end{equation}
 where $\alpha (t, y)$ and $\beta (t, y)$ are any solutions of the adjoint equation (\ref{anh-1:eq24})
 to the linear heat equation (\ref{anh-1:eq25}).\\[1ex]
 \textbf{Proposition 2.6.}
 The equation
 \begin{equation}
 \label{anh-int:eq27}
  u_t = (f(u) u_x)_x + k u_{yy} - \delta^2 {\cal F}(u), \quad f(u) = {\cal F}'(u),
 \end{equation}
  satisfies the nonlinear self-adjointness condition
 (\ref{anh-int.2:eq.6}) with the substitution (\ref{anh-1:eq3}) of the form
 \begin{equation}
 \label{anh-1:eq26}
 v = \alpha (t, y) {\rm e}^{\delta x} + \beta (t, y) {\rm e}^{- \delta x},
 \end{equation}
 where $\alpha (t, y)$ and $\beta (t, y)$ are any solutions of the adjoint equation (\ref{anh-1:eq24})
 to the linear heat equation (\ref{anh-1:eq25}).

 \section{Conservation laws}
 \label{anh-2}
 \setcounter{equation}{0}

 \subsection{Computation of conserved vectors for Equation (\ref{anh-int:eq1})}
  \label{anh-2.1}

 Here we construct the conserved vector (\ref{anh-int.3:eq.2}) for Eq. (\ref{anh-int:eq1}),
 \begin{equation}
 \label{anh-2:eq1}
  u_t = f(u) u_{xx} +g(u)u_{yy} +h(u)u_{zz}+ f'(u) u_x^2 +g'(u) u_y^2 +h'(u)
   u_z^2,
 \end{equation}
 associated with its \textit{translational symmetries}. We specify the notation by writing the
 symmetry generator (\ref{anh-int.3:eq.1}) in the form
 \begin{equation}
 \label{anh-2:eq2}
   X= \xi^1\frac{\partial}{\partial t}+ \xi^2\frac{\partial}{\partial x} + \xi^3\frac{\partial}{\partial y}
    + \xi^4\frac{\partial}{\partial z}+ \eta\frac{\partial}{\partial
    u}\,\cdot
 \end{equation}
 Then the expression (\ref{anh-int.3:eq.4}) becomes
 \begin{equation}
 \label{anh-2:eq3}
 W=\eta-\xi^1 u_t- \xi^2 u_x- \xi^3 u_y- \xi^4 u_z
 \end{equation}
 and the conservation equation (\ref{anh-int.3:eq.3}) means that
 the following equation holds on the solutions of Eq. (\ref{anh-2:eq1}):
 \begin{equation}
 \label{anh-2:eq4}
  D_t \big(C^1\big)+ D_x \big(C^2\big) +D_y \big(C^3\big)+D_z \big(C^4\big)=0.
 \end{equation}

The formal Lagrangian (\ref{anh-int.3:eq.5}) for Eq.
(\ref{anh-2:eq1}) is
 \begin{equation}
 \label{anh-2:eq5}
   {\mathcal L}= v[f(u) u_{xx} +g(u)u_{yy} +h(u)u_{zz}+ f'(u) u_x^2 +g'(u) u_y^2 +h'(u)
   u_z^2-u_t].
 \end{equation}
 Due to the specific dependence of the formal Lagrangian
 (\ref{anh-2:eq5}) on the derivatives $u_i, u_{ij},$ the components of the vector (\ref{anh-int.3:eq.2})
 are written
   \begin{align}
  C^1 &= W\frac{\partial \mathcal L}{\partial u_t}\,,\notag\\[1ex]
 C^2 &=W\left[\frac{\partial \mathcal L}{\partial u_x}-D_x \left( \frac{\partial \mathcal L}{\partial u_{xx}}\right) \right]+
     D_x(W) \frac{\partial \mathcal L}{\partial u_{xx}}\,,  \notag\\[1ex]
   C^3 &=W\left[\frac{\partial \mathcal L}{\partial u_y}-D_y \left( \frac{\partial \mathcal L}{\partial u_{yy}}\right) \right]+
     D_y(W) \frac{\partial \mathcal L}{\partial u_{yy}}\,,  \notag\\[1ex]
     C^4 &=W\left[\frac{\partial \mathcal L}{\partial u_z}-D_z \left( \frac{\partial \mathcal L}{\partial u_{zz}}\right) \right]+
     D_z(W) \frac{\partial \mathcal L}{\partial u_{zz}}\,\cdot  \notag
   \end{align}
 Substituting here the explicit expression (\ref{anh-2:eq5}) of ${\mathcal
 L}$ we obtain:
   \begin{align}
  C^1 &= -Wv,\notag\\[1ex]
 C^2 &= W\left[f'(u) u_x v-f(u) v_x \right]+
     f(u) v D_x(W),\notag\\[1ex]
   C^3 &= W\left[g'(u) u_y v-g(u) v_y \right]+
     g(u) v D_y(W),  \label{anh-2:eq6}\\[1ex]
     C^4 &= W\left[h'(u) u_z v-h(u) v_z \right]+
     h(u) v D_z(W). \notag
   \end{align}

 Eqs. (\ref{anh-int:eq1}) and (\ref{anh-int:eq2}) with arbitrary coefficients $f(u), g(u),
 h(u)$ are invariant  under the groups of translations of $t, x, y, z$
 with the generators
    \begin{equation}
\label{anh-2:eq7}
 X_1=\frac{\partial}{\partial t}\,, \quad  X_2=\frac{\partial}{\partial x}\,,\quad X_3=\frac{\partial}{\partial y}\,, \quad X_4=\frac{\partial}{\partial z}\,\cdot
  \end{equation}
 Eq. (\ref{anh-int:eq1}) has also a dilation symmetry. Moreover, both equations (\ref{anh-int:eq1}) and (\ref{anh-int:eq2}) may have more symmetries in  certain particular
 cases \cite{ibr94}, but we don't consider them here.

  Let us apply the formula (\ref{anh-2:eq6}) to the symmetry $X_2.$
 The corresponding quantity (\ref{anh-2:eq3}) equals
 $W = -u_x.$ We substitute it in (\ref{anh-2:eq6}) and obtain
 \begin{align}
  C^1 &= v u_x,\notag\\[1ex]
 C^2 &= - f'(u) v u_x^2 +f(u) u_x v_x -
     f(u) v u_{xx},\notag\\[1ex]
   C^3 &= - g'(u) v u_x u_y +g(u) u_x v_y -
     g(u) v u_{xy},  \label{anh-2:eq8}\\[1ex]
     C^4 &= - h'(u) v u_x u_z + h(u) u_x v_z -
     h(u) v u_{xz}. \notag
  \end{align}
 We have to substitute here the expression (\ref{anh-1:eq8}) for $v,$
  \begin{equation}
 \tag{\ref{anh-1:eq8}}
 v = a_1\,xyz + a_2\,xy + a_3\,xz + a_4\,y z + a_5\,x + a_6\,y + a_7\,z +
 a_8.
 \end{equation}
 Since $v$ is a given function whereas $u$ is any solution of Eq.
 (\ref{anh-2:eq1}), we want to simplify the conserved
 vector (\ref{anh-2:eq8}) by transforming it in an equivalent
 conserved vector which conserved density $C^1$ contains $u$ instead
 of $u_x.$ To this end, we use the identity $v u_x = D_x(uv) - u
 v_x$ and write $C^1$ in (\ref{anh-2:eq8}) in the form
 $$
 C^1 = \widetilde C^1 + D_x(uv),
 $$
 where
 \begin{equation}
 \label{anh-2:eq9}
 \widetilde C^1 =  - u v_x.
 \end{equation}
 Then we transfer the term $D_x(uv)$ from $C^1$ to $C^2$ using the usual procedure (see, e.g.
 \cite{ibr10a}, Section 8.1). Namely, since the total differentiations
 commute with each other, we have
 $$
 D_t\big(\widetilde C^1 + D_x(uv)\big) + D_x \big(C^2\big) = D_t\big(\widetilde C^1\big) + D_x \big(C^2  +
 D_t(uv)\big).
 $$
 Therefore the conservation equation (\ref{anh-2:eq4}) for the vector
 (\ref{anh-2:eq8}) can be equivalently rewritten  in the form
 \begin{equation}
 \label{anh-2:eq10}
  D_t \big(\widetilde C^1\big)+ D_x \big(\widetilde C^2\big) +D_y \big(C^3\big)+D_z
  \big(C^4\big)=0,
 \end{equation}
 where $\widetilde C^1$ has the form (\ref{anh-2:eq9}) and $\widetilde C^2$ is
 given by
 $$
 \widetilde C^2 = C^2 +  D_t(uv).
 $$
 Let us work out the above expression for $\widetilde C^2.$ Invoking that $D_t(v) = 0$ due to Eq. (\ref{anh-1:eq8}),
 and using Eqs. (\ref{anh-int:eq1}), (\ref{anh-2:eq1}) we have
 $$
 \widetilde C^2 = C^2 + v u_t = C^2 + v \big[f(u) u_{xx} + f'(u) u_x^2 +
  D_y \big(g(u)u_y\big) +  D_z \big(h(u)u_z\big)\big].
 $$
We substitute here the expression of $C^2$ from Eqs.
(\ref{anh-2:eq8}) and obtain:
 $$
 \widetilde C^2 = f(u) u_x v_x +
  v D_y \big(g(u)u_y\big) + v D_z \big(h(u)u_z\big).
 $$
 We simplify the latter expression for $\widetilde C^2$  by
 noting that
 $$
 v D_y \big(g(u)u_y\big) = D_y \big(g(u) v u_y\big) - g(u) u_y v_y,
 $$
 $$
  v D_z \big(h(u)u_z\big) = D_z \big(h(u) v u_z\big) - h(u) u_z
 v_z
 $$
 Therefore we transfer the terms $D_y \big(g(u) v u_y\big)$ and $D_z \big(h(u) v u_z\big)$
 to $C^3$ and $C^4,$ respectively, and obtain:
 \begin{equation}
 \label{anh-2:eq11}
 \widetilde C^2 = f(u) u_x v_x - g(u) u_y v_y - h(u) u_z.
 \end{equation}
 Now the components $C^3$ and $C^4$ of the vector (\ref{anh-2:eq8}) become:
 $$
 \widetilde C^3 = C^3 +  D_x \big(g(u) v u_y\big), \quad  \widetilde C^4 = C^4 +  D_x \big(h(u) v u_z\big).
 $$
 After substituting here the expressions of $C^3$ and $C^4$ from
 (\ref{anh-2:eq8}) we have
 $$
 \widetilde C^3 = g(u) \big(u_x v_y + u_y v_x\big), \quad  \widetilde C^4 =
 h(u) \big(u_x v_z + u_z v_x\big).
 $$
 Combining these expressions with (\ref{anh-2:eq9}),
 (\ref{anh-2:eq11})  and ignoring the tilde, we arrive at the following conserved vector
 which is equivalent to (\ref{anh-2:eq8}):
 \begin{align}
  C^1 &= -uv_x,\notag\\[1ex]
 C^2 &= f(u)u_xv_x -g(u)u_yv_y-h(u)u_zv_z,   \label{anh-2:eq12}\\[1ex]
   C^3 &=g(u)(u_xv_y+ u_yv_x),  \notag\\[1ex]
      C^4 &=h(u)(u_xv_z+ u_zv_x). \notag
   \end{align}
The vector (\ref{anh-2:eq12}) involves the first-order derivatives
of the variable $v$ given by Eq. (\ref{anh-1:eq8}). Therefore the
vector (\ref{anh-2:eq12}) contains seven parameters $a_1, \ldots,
a_7.$ In fact, it is a linear combination of \textit{seven} linearly
independent conserved vectors obtained from (\ref{anh-2:eq12})
setting by turns one of the parameters  $a_i$ equal to 1 and the
others equal to 0. But some of these seven vectors are trivial in
the sense that their divergence is identically zero, i.e.  the
conservation equation (\ref{anh-2:eq4}) is satisfied identically.
For example, setting in (\ref{anh-2:eq12}) $a_6 = 1, a_1= \cdots =
a_7 = 0,$ i.e. $v = y,$ we obtain the vector
 $$
 C^1 = 0, \quad C^2 = - g(u) u_y, \quad C^3 = g(u) u_x, \quad C^4 = 0.
 $$
For this vector Eq. (\ref{anh-2:eq4}) is satisfied identically,
 $$
D_t(C^1) +  D_x(C^2)+D_y(C^3) +D_z(C^4)= - D_x\big(g(u) u_y\big) +
D_y\big(g(u) u_x\big) \equiv 0.
 $$

 Let us single out the \textit{nontrivial conserved vectors}. Since $v$ given by Eq.
(\ref{anh-1:eq8}),
  the conservation equation (\ref{anh-2:eq4}) for the
vector (\ref{anh-2:eq12}) is written as
 \begin{equation}
 \label{anh-2:eq13}
     D_t(C^1) +  D_x(C^2)+D_y(C^3) +D_z(C^4)=v_xF,
  \end{equation}
where $F$ is given by Eq.~(\ref{anh-1:eq1}),
 $$
 F = - u_t + (f(u)
u_x)_x + (g(u) u_y)_y + (h(u) u_z)_z.
 $$
 Then, specifying the expression of $v_x$ from (\ref{anh-1:eq8}), we write (\ref{anh-2:eq13}) in the
  form
\begin{equation}
\label{anh-2:eq14}
 D_t(C^1) +  D_x(C^2)+D_y(C^3) +D_z(C^4)=(a_1 yz + a_2 y + a_3 z + a_5)
 F.
  \end{equation}
 Eq. (\ref{anh-2:eq14}) shows that we have only \textit{four} nontrivial conserved vectors. They correspond to
 $a_1, a_2, a_3$ and $a_5,$ i.e. they are obtained from
 (\ref{anh-2:eq12}) setting by turns one of these four parameters to
 be equal to 1, the others  equal to 0. For example, the nontrivial
 conserved vector (\ref{anh-2:eq12}) corresponding to $a_5$ is
\begin{equation}
\label{anh-2:eq15}
 C^1 = - u, \quad C^2 = f(u) u_x, \quad C^3 = g(u) u_y, \quad
 C^4 = h(u) u_z.
  \end{equation}
 The conservation equation (\ref{anh-2:eq4})  for the vector (\ref{anh-2:eq15}) coincides
 with Eq. (\ref{anh-int:eq1}).

 Thus, the nontrivial conserved vectors are obtained by substituting
 in (\ref{anh-2:eq12}) the expression
 (\ref{anh-1:eq8}) for $v$ with $a_4 = a_6 = a_7 = 0.$ The resulting
 vector
 \begin{align}
  C^1 &= -(a_1yz + a_2 y+a_3 z+ a_5)u,\notag\\[1ex]
 C^2 &= (a_1yz + a_2 y+a_3 z+ a_5)f(u) u_x \label{anh-2:eq16}\\[.1ex]
 & - (a_1z +a_2)x g(u) u_y
 - (a_1y+a_3) x h(u) u_z,  \notag\\[1ex]
   C^3 &= (a_1z +a_2) g(u) (x u_x + y u_y) + (a_3 z + a_5) g(u) u_y,  \notag\\[1ex]
      C^4 &= (a_1y+a_3) h(u) (x u_x+ z u_z) + (a_2 y + a_5) h(u) u_z \notag
   \end{align}
   is the linear combination with the coefficients $a_5, a_1, a_2, a_3$ of four linearly independent  vectors,
   namely the vector (\ref{anh-2:eq15}) and the following three vectors:
 \begin{align}
 \label{anh-2:eq17}
 & C^1 = - yz u,\quad
 C^2 = yz f(u) u_x
  - x z  g(u) u_y
 - xy h(u) u_z,  \notag\\[1ex]
  & C^3 = z g(u) (x u_x + y u_y),  \quad
      C^4 = y h(u)(x u_x+ z u_z);
   \end{align}
 \begin{align}
 \label{anh-2:eq18}
 & C^1 = - y u,\quad
 C^2 =  y f(u) u_x -  x g(u) u_y, \notag\\[1ex]
  & C^3 = g(u) (x u_x + y u_y),  \quad
      C^4 = y h(u) u_z;
   \end{align}
   \begin{align}
     \label{anh-2:eq19}
 & C^1 = - z u,\quad
 C^2 =  z f(u) u_x -  x h(u) u_z, \notag\\[1ex]
  & C^3 = z g(u) u_y,  \quad
      C^4 = h(u)  (x u_x + z u_z).
   \end{align}

 The conserved vectors associated with the symmetry $X_3$ from
 (\ref{anh-2:eq7}) can be obtained from the above results merely by the
  permutation $x\leftrightarrow y,$
 followed by the permutations  $f\leftrightarrow g$ and $C^2\leftrightarrow C^3.$
 This procedure maps the vector (\ref{anh-2:eq12}) to the following
 conserved vector:
  \begin{align}
  C^1 &= -uv_y,\notag\\[1ex]
 C^2 &= f(u)(u_yv_x+ u_xv_y),   \label{anh-2:eq20}\\[1ex]
   C^3 &= g(u)u_yv_y -f(u)u_xv_x-h(u)u_zv_z,  \notag\\[1ex]
      C^4 &=h(u)(u_yv_z+ u_zv_y).  \notag
   \end{align}
   Accordingly, Eq. (\ref{anh-2:eq14}) becomes  the
   following conservation equation for the vector
   (\ref{anh-2:eq20}):
 \begin{equation}
 \label{anh-2:eq21}
 D_t(C^1) +  D_x(C^2)+D_y(C^3) +D_z(C^4)=(a_1 xz + a_2 x + a_4 z + a_6)
 F.
  \end{equation}
 It shows that the nontrivial conserved vectors are obtained by substituting
 in (\ref{anh-2:eq20}) the expression
 (\ref{anh-1:eq8}) for $v$ with $a_3 = a_5 = a_7 = 0.$ The resulting
 vector
 \begin{align}
  C^1 &= -(a_1 xz + a_2 x + a_4 z + a_6)u,\notag\\[1ex]
     C^2 &= (a_1z +a_2) f(u) (x u_x + y u_y) + (a_4 z + a_6) f(u) u_x,\label{anh-2:eq22}\\[1ex]
 C^3 &= (a_1 xz + a_2 x + a_4 z + a_6)g(u) u_y \notag\\[.1ex]
 & - (a_1z +a_2)y f(u) u_x
 - (a_1x+a_4) y h(u) u_z,  \notag\\[1ex]
      C^4 &= (a_1 x+a_4) h(u) (y u_y+ z u_z) + (a_2 x + a_6) h(u) u_z \notag
   \end{align}
 is the linear combination with the coefficients $a_6, a_1, a_2, a_4$ of four linearly independent  vectors,
   namely the vector (\ref{anh-2:eq15}) and the following three vectors:
 \begin{align}
 \label{anh-2:eq23}
 & C^1 = - xz u,\quad
 C^2 = z f(u) (x u_x + y u_y),  \notag\\[1ex]
  & C^3 = xz g(u) u_y
  - y z  f(u) u_x
 - xy h(u) u_z,  \\[1ex]
  & C^4 = x h(u)(y u_y+ z u_z);\notag
   \end{align}
 \begin{align}
 \label{anh-2:eq24}
 & C^1 = - x u,\quad
 C^2 = f(u) (x u_x + y u_y), \notag\\[1ex]
  & C^3 = x g(u) u_y -  y f(u) u_x,  \quad
      C^4 = x h(u) u_z;
   \end{align}
   \begin{align}
     \label{anh-2:eq25}
 & C^1 = - z u,\quad
 C^2 = z f(u) u_x, \\[1ex]
  & C^3 = z g(u) u_y -  y h(u) u_z,  \quad
      C^4 = h(u)  (y u_y + z u_z).\notag
   \end{align}

Proceeding as above with the symmetry $X_4$ from
 (\ref{anh-2:eq7}) we obtain, in addition to (\ref{anh-2:eq15}), (\ref{anh-2:eq17})-(\ref{anh-2:eq19}) and
 (\ref{anh-2:eq23})-(\ref{anh-2:eq25}), the following conserved
vectors:
 \begin{align}
 \label{anh-2:eq26}
 & C^1 = - xy u,\quad
 C^2 = y f(u) (x u_x + z u_z),  \notag\\[1ex]
  & C^3 = x g(u)(y u_y+ z u_z),  \\[1ex]
  & C^4 = xy h(u) u_z - y z  f(u) u_x
 - xz g(u) u_y;\notag
   \end{align}
 \begin{align}
 \label{anh-2:eq27}
 & C^1 = - x u,\quad
 C^2 = f(u) (x u_x + z u_z), \notag\\[1ex]
  & C^3 = x g(u) u_y,  \quad
      C^4 = x h(u) u_z  -  z f(u) u_x;
   \end{align}
   \begin{align}
     \label{anh-2:eq28}
 & C^1 = - y u,\quad
 C^2 = y f(u) u_x, \\[1ex]
  & C^3 = g(u)  (y u_y + z u_z),  \quad
      C^4 = y h(u) u_z -  z g(u) u_y.\notag
   \end{align}

Finally, we turn to the time-translational symmetry $X_1$ from
(\ref{anh-2:eq7}). In this case $W=-u_t.$ Replacing $u_t$ by the
right-hand side of Eq. (\ref{anh-int:eq1})  we obtain from the first
equation (\ref{anh-2:eq6}):
 \begin{equation}
 \label{anh-2:eq29}
  C^1  = v\left[ D_x (f(u) u_x) + D_y(g(u) u_y)
 +D_z(h (u) u_z)\right].
   \end{equation}
   Now we observe that
 $$
 v  D_x (f(u) u_x)=   D_x \left[v f(u) u_x-v_x {\cal F} (u)\right],
 $$
 where we denote ${\cal F} (u) = \int f(u)du $ and use the equation $v_{xx} = 0$ resulting from the
 representation (\ref{anh-1:eq8}) of $v.$ Transforming likewise
 two other terms in (\ref{anh-2:eq29}) we write $C^1$ in the divergent form:
 $$
 C^1  = D_x \left[vf(u) u_x-v_x {\cal F} (u)\right]+D_y \left[vg(u) u_y-v_y {\cal G}(u)\right]+ D_z \left[vh(u) u_z-v_z {\cal H}(u)\right],
 $$
where ${\cal G}(u)=\int g(u) du,$ \ ${\cal H}(u)=\int h(u) du.$ Now
we can transfer all terms of $C^1$ to the components $C^2, C^3, C^4$
and obtain $C^1 = 0.$ The calculation shows that after this transfer
we will have $C^1 = C^2 = C^3 = C^4 = 0.$ Hence, $X_1$ does not lead
to a non-trivial conservation law.

Thus, we have proved the following statement.\\[1ex]
 \textbf{Theorem 3.1.} The translational symmetries (\ref{anh-2:eq7}) of  Eq. (\ref{anh-2:eq1}) with arbitrary
 coefficients $f(u), g(u), h(u)$ provide ten linearly independent
 conserved vectors (\ref{anh-2:eq15}), (\ref{anh-2:eq17})-(\ref{anh-2:eq19}) and
 (\ref{anh-2:eq23})-(\ref{anh-2:eq28}).\\[1ex]
 \textbf{Remark 3.1.}  Eq. (\ref{anh-1:eq22}) is nonlinearly
 self-adjoint with the substitution (\ref{anh-1:eq26}) containing
 arbitrary solutions $\alpha (t, y),$ $\beta (t, y),$ $\gamma(t, y),$
 $\sigma(t, y)$ of the adjoint equation (\ref{anh-1:eq24}) to the
 one-dimensional linear heat equation. Therefore the conserved vector
 constructed by the above procedure for Eq. (\ref{anh-1:eq22}) will
 contain arbitrary solutions of Eq. (\ref{anh-1:eq24}).

 \subsection{Conserved vectors for Equation (\ref{anh-int:eq18})}
  \label{anh-2.2}

 As mentioned in Section \ref{anh-int.1}, the anisotropic heat equation (\ref{anh-int:eq2})
 with an external source $q(u) \not= 0$ does not have a conservation
 form.  However, Eq. (\ref{anh-int:eq2}) can be rewritten in a conservation form
 $$
  D_t \big(C^1\big)+ D_x \big(C^2\big) +D_y \big(C^3\big)=0
 $$
 if it is nonlinearly self-adjoint, for example, in the special cases
 (\ref{anh-int:eq18}) and (\ref{anh-int:eq19}). We will
find here the conservation form for Eq. (\ref{anh-int:eq18}). The
calculations are similar for Eq. (\ref{anh-int:eq19}).

We write Eq. (\ref{anh-int:eq18}) in the form
 \begin{equation}
 \label{anh-2:eq30}
  u_t = f(u) u_{xx} +g(u)u_{yy} + f'(u) u_x^2 +g'(u) u_y^2  + \omega^2 {\cal F}(u), \quad \omega =
  {\rm const.,}
 \end{equation}
and have the formal Lagrangian
 \begin{equation}
 \label{anh-2:eq31}
   {\mathcal L}= v[f(u) u_{xx} +g(u)u_{yy} + f'(u) u_x^2 +g'(u) u_y^2  + \omega^2 {\cal F}(u)-u_t].
 \end{equation}
 For this formal Lagrangian
 Eqs. (\ref{anh-int.3:eq.2}) yield (cf. Eqs. (\ref{anh-2:eq6}))
   \begin{align}
    \label{anh-2:eq32}
  C^1 &= -Wv,\notag\\[1ex]
 C^2 &= W\left[f'(u) u_x v-f(u) v_x \right]+
     f(u) v D_x(W),\\[1ex]
   C^3 &= W\left[g'(u) u_y v-g(u) v_y \right]+
     g(u) v D_y(W). \notag
   \end{align}

 Eq. (\ref{anh-2:eq30}) admits the three-dimensional Lie algebra
 spanned by the operators $X_1, X_2, X_3$ from (\ref{anh-2:eq7}).
  Let us apply the formula (\ref{anh-2:eq32}) to the symmetry $X_2.$
 In this case $W = -u_x$  and (\ref{anh-2:eq32}) is written (see Eqs. (\ref{anh-2:eq8}))
 \begin{align}
  \label{anh-2:eq33}
  C^1 &= v u_x,\notag\\[1ex]
 C^2 &= - f'(u) v u_x^2 +f(u) u_x v_x -
     f(u) v u_{xx},\\[1ex]
   C^3 &= - g'(u) v u_x u_y +g(u) u_x v_y -
     g(u) v u_{xy},\notag
  \end{align}
 where $v$ should be replaced by its expression (\ref{anh-1:eq20}),
 \begin{equation}
 \tag{\ref{anh-1:eq20}}
 v = \left(A_1 y + B_1\right) \cos (\omega x) + \left(A_2 y + B_2\right) \sin (\omega
 x).
 \end{equation}
 Let us simplify the vector (\ref{anh-2:eq33}) in the same way  as in Section
 \ref{anh-2.2}. We write
$$
 C^1 = \widetilde C^1 + D_x(uv),
 $$
 where
 \begin{equation}
 \label{anh-2:eq34}
 \widetilde C^1 =  - u v_x,
 \end{equation}
 and replace $C^2$ by
 $$
 \widetilde C^2 = C^2 +  D_t(uv).
 $$
 Hence
\begin{align}
 \widetilde C^2 & = C^2 + v u_t\notag\\
 & = C^2 + v \big[f(u) u_{xx} + f'(u) u_x^2 + D_y\big(g(u)u_y\big)  + \omega^2 {\cal
 F}(u)\big].\notag
 \end{align}
The substitution of the expression (\ref{anh-2:eq32}) for $C^2$
yields:
 $$
 \widetilde C^2  = f(u) u_x v_x + \omega^2 {\cal F}(u)
  + v D_y \big(g(u)u_y\big).
 $$
 One  can verify that the following equation holds:
 $$
 v D_y \big(g(u)u_y\big) =  D_y \left[v g(u)u_y - {\cal G} (u)
 v_y\right].
 $$
 It is obtained by introducing the function ${\cal G} (u) = \int g(u)du $ and noting that the equation $v_{yy} = 0$ is valid
 for the
 representation (\ref{anh-1:eq20}) of $v.$ Hence $\widetilde C^2$ can be reduced to
 \begin{equation}
 \label{anh-2:eq35}
 \widetilde C^2 = f(u) u_x v_x + \omega^2 {\cal F}(u),
 \end{equation}
 whereas $C^3$ becomes
  $$
 \widetilde C^3 = C^3 + D_x\left[v g(u)u_y - {\cal G} (u) v_y\right].
 $$
 The latter equation upon inserting the expression (\ref{anh-2:eq32}) for
 $C^3$ yields:
 \begin{equation}
 \label{anh-2:eq36}
 \widetilde C^3 = g(u) u_y v_x -  {\cal G} (u) v_{xy}.
 \end{equation}

 Collecting Eqs. (\ref{anh-2:eq34})-(\ref{anh-2:eq36}) and ignoring the tilde we arrive at the vector
 \begin{align}
 \label{anh-2:eq37}
  C^1 &= - u v_x,\notag\\[1ex]
 C^2 &= f(u) u_x v_x + \omega^2 {\cal F}(u)v ,\\[1ex]
   C^3 &= g(u) u_y v_x -  {\cal G} (u) v_{xy},\notag
 \end{align}
 where $v$ should be replaced by its expression (\ref{anh-1:eq20}).
 The vector (\ref{anh-2:eq37}) satisfies the conservation equation in the following
 form:
 $$
  D_t \big(C^1\big)+ D_x \big(C^2\big) +D_y \big(C^3\big)=v_x \left[(f(u) u_x)_x + (g(u) u_y)_y + \omega^2 {\cal F}(u) - u_t\right].
 $$

 The expression (\ref{anh-1:eq20}) for $v$ contains four arbitrary constants $A_1, A_2, B_1, B_2.$
 Accordingly, the vector (\ref{anh-2:eq37}) is a linear combination
 of four linearly independent vectors. Hence, we have demonstrated
 the following statement.\\[1ex]
 \textbf{Theorem 3.2.} The invariance of Eq. (\ref{anh-2:eq18}) with respect
 to the one-parameter group of translations of $x$ with the generator $X_2$ provides the following four linearly independent
 conserved vectors:
 \begin{align}
 \label{anh-2:eq38}
 & C^1 =  \sin(\omega x)\, u,\quad
 C^2 = - \sin(\omega x)\, f(u) u_x + \omega \cos(\omega x)\, {\cal F}(u),\notag\\[.5ex]
 & C^3 = - \sin(\omega x)\, g(u)u_y;
   \end{align}
 \begin{align}
 \label{anh-2:eq39}
 & C^1 =  \cos(\omega x)\, u,\quad
 C^2 = - \cos(\omega x) f(u) u_x - \omega \sin(\omega x)\,  {\cal F}(u),\notag\\[.5ex]
 & C^3 = - \cos(\omega x)\, g(u)u_y;
   \end{align}
 \begin{align}
 \label{anh-2:eq40}
 & C^1 = y \sin(\omega x)\, u,\quad
 C^2 = - y \sin(\omega x)\, f(u) u_x + \omega y \cos(\omega x)\, {\cal F}(u), \notag\\[.5ex]
  & C^3 = - y \sin(\omega x)\, g(u)u_y +  \sin(\omega x) {\cal G}(u);
   \end{align}
  \begin{align}
  \label{anh-2:eq41}
 & C^1 = y \cos(\omega x)\, u,\quad
 C^2 = - y \cos(\omega x)\, f(u) u_x - \omega y \sin(\omega x)\, {\cal F}(u),\notag \\[.5ex]
  & C^3 = - y \cos(\omega x)\, g(u)u_y +  \cos(\omega x) {\cal G}(u).
   \end{align}
 \textbf{Remark 3.2.} The conserved vectors provided by the generator $X_3$ of
  the group of translations of $y$ can be computed likewise.
  The generator $X_1$ of the time-translations provides only the
  trivial conserved vector.\\[1ex]
 \textbf{Remark 3.3.} Using the conserved vectors (\ref{anh-2:eq38})-(\ref{anh-2:eq41})
  one can write Eq. (\ref{anh-int:eq18}) in four different conservation
  forms. For example, the  vector (\ref{anh-2:eq38}) satisfies the conservation equation
 $$
  D_t \big(C^1\big)+ D_x \big(C^2\big) +D_y \big(C^3\big)
  = \sin(\omega x)\, \left[u_t - (f(u) u_x)_x - (g(u) u_y)_y - \omega^2 {\cal F}(u)\right].
 $$
Accordingly, Eq. (\ref{anh-int:eq18}) can be replaced by the
following
 conservation equation:
 \begin{align}
 \label{anh-2:eq42}
  D_t \left[\sin(\omega x)\, u\right]
  &- D_x \left[\sin(\omega x)\, f(u) u_x + \omega \cos(\omega x)\, {\cal F}(u)\right]\notag\\[.5ex]
 & - D_y \left[\sin(\omega x)\, g(u)u_y\right] = 0.
   \end{align}

\end{document}